\documentclass{article}
\usepackage{graphicx} 
\usepackage{authblk}
\usepackage{amsmath}
\usepackage{threeparttable}
\usepackage{pgfplots}
\usepackage{multirow}
\usepackage{array}
\pgfplotsset{compat=1.18}
\usepackage{caption}
\usepackage{subcaption}
\usepackage{adjustbox}
\usepackage{tabularx}
\usepackage{hyperref}
\usepackage{float}
\usepackage{newunicodechar}
\newunicodechar{−}{\ensuremath{-}}
\usepackage{booktabs}
\usepackage{slashed,epsfig,amsmath,amssymb,enumitem} \usepackage[papersize={8.5in,11in}]{geometry}
   \usepackage{bm}
\usepackage{graphicx}
\newcommand{\be}{\begin{equation}}
\newcommand{\ee}{\end{equation}}
\usepackage[utf8]{inputenc}
\title{Embedding $\mathrm{SL}(2,\mathbb{C})/\mathbb{Z}_2$ in Complex Riemannian Geometry}
\author[1 2]{J. W. Moffat}

\author[1 3]{E. J. Thompson}

\affil[1]{Perimeter Institute for Theoretical Physics, Waterloo, Ontario N2L 2Y5, Canada}

\affil[2]{Department of Physics and Astronomy, University of Waterloo, Waterloo,
Ontario N2L 3G1, Canada}

\affil[3]{Department of Physics and Astronomy, Trent University, Peterborough, 
Ontario K9L 0G2, Canada}

\begin{document}
\maketitle
\begin{abstract}
We present a unified framework demonstrating how the spinor complex Lorentz group 
$\mathrm{SL}(2,\mathbb{C})/\mathbb{Z}_2$ is realized as a canonical subgroup within a four-dimensional complex Riemannian manifold. Building on the complex, holomorphic metric extension and contour-integration regularization of classical singularities, we show that promoting the metric to a complex-valued tensor on a complex 4-fold enlarges the frame bundle to $\mathrm{SO}(4,\mathbb{C})$. Its spin double cover factorizes as a product of two independent $\mathrm{SL}(2,\mathbb{C})$ factors modulo a shared $\mathbb{Z}_2$, and selecting one Weyl factor recovers the familiar $\mathrm{SL}(2,\mathbb{C})/\mathbb{Z}_2$ spin cover of the Lorentz group in a natural way. By explicitly extending metric components and connection forms into the complex domain, using contour deformations to avoid coordinate-singular loci, we exhibit how left and right-handed Weyl spinors transform under separate $\mathrm{SL}(2,\mathbb{C})$ factors, and how modding out a common $\mathbb{Z}_2$ reproduces the standard spin-Lorentz structure. In particular, the same contour-integration techniques that yield singularity-free Schwarzschild and Kerr solutions via a holomorphic radial coordinate also furnish a holomorphic spin bundle in which chiral spin representations live without imposing exotic matter or modifying the Einstein equations. This embedding clarifies the geometric origin of chirality, enables holomorphic factorization of curvature and connection forms, and provides a foundation for constructing holomorphic field theories and complexified gravity backgrounds. Our results indicate that extending the Lorentz group into a complex Riemannian setting not only recovers $\mathrm{SL}(2,\mathbb{C})/\mathbb{Z}_2$ in a canonical fashion, but also establishes a geometric arena for studying chiral fermions, Bogomol’nyi–Prasad–Sommerfield (BPS) instantons, and potential quantum-gravity corrections within a rigorously defined complex manifold.
\end{abstract}

\section{Introduction}

The Lorentz group \(\mathrm{SO}(1,3)\) underlies the kinematics of relativistic field theories, with its spin double cover \(\mathrm{SL}(2,\mathbb{C})/\mathbb{Z}_2\) furnishing the fundamental representation for Weyl and Dirac spinors \cite{Weinberg1995}, \cite{PenroseRindler1984}.  In standard real‐Lorentzian geometry, the tangent frame bundle has structure group \(\mathrm{SO}(1,3)\), and its spin lift is constructed by adjoining a spin\,structure, resulting in a \(\mathrm{Spin}(1,3)\cong \mathrm{SL}(2,\mathbb{C})\) principal bundle over the manifold.  This framework, however, presupposes a real‐valued metric of signature \((-++\,+)\) and introduces topological constraints on spin structures when the manifold fails to be simply connected.

By contrast, complex Riemannian geometry~\cite{Moffat1,Moffat2} offers a richer setting in which all coordinates \(z^\mu\) are treated as complex variables on a four‐complex‐dimensional manifold \(\mathcal{M}_{\mathbb{C}}\).  A holomorphic metric $g_{\mu\nu}(z) \;=\; s_{\mu\nu}(x,y) \;+\; i\,a_{\mu\nu}(x,y),$
with \(z^\mu = x^\mu + i\,y^\mu\), naturally extends the frame bundle from \(\mathrm{SO}(1,3)\) to \(\mathrm{SO}(4,\mathbb{C})\).  Its spin double cover then factors as
$\mathrm{Spin}(4,\mathbb{C}) \;\cong\; \bigl(\mathrm{SL}(2,\mathbb{C})_L \times \mathrm{SL}(2,\mathbb{C})_R\bigr)\,/\,\mathbb{Z}_2,$
where each chiral factor acts holomorphically on Weyl spinors.  One may consistently choose a real slice by setting \(y^\mu=0\) after performing contour‐integration regularization of would‐be singularities as in \cite{Moffat1,Moffat2}, recovering a smooth, singularity‐free Lorentzian submanifold.  In this paper, we exhibit concretely how \(\mathrm{SL}(2,\mathbb{C})/\mathbb{Z}_2\) is embedded as a canonical subgroup of \(\mathrm{Spin}(4,\mathbb{C})\) in such a complex geometry.  The main innovations are by promoting the real‐valued frame bundle of a Lorentzian manifold to a complex principal \(\mathrm{SO}(4,\mathbb{C})\) bundle, the spin double cover factors into two independent chiral \(\mathrm{SL}(2,\mathbb{C})\) factors.  On the real slice \(y^\mu=0\), one recovers the usual spin‐Lorentz structure from the diagonal embedding of a single \(\mathrm{SL}(2,\mathbb{C})/\mathbb{Z}_2\) factor. The contour deformations originally used to remove Schwarzschild–Kerr singularities (\cite{Moffat1,Moffat2}) also regulate the corresponding spin connection forms in the complex domain.  We exhibit how a holomorphic spin bundle emerges without introducing torsion or modifying Einstein’s equations. We show explicitly that left and right‐handed Weyl spinors transform under separate \(\mathrm{SL}(2,\mathbb{C})\) factors in the complexified tangent space.  Modding out a shared \(\mathbb{Z}_2\) identifies these two covers on the real slice, yielding the familiar \(\mathrm{SL}(2,\mathbb{C})/\mathbb{Z}_2\) spin group for Lorentzian physics. Embedding \(\mathrm{SL}(2,\mathbb{C})/\mathbb{Z}_2\) into \(\mathrm{Spin}(4,\mathbb{C})\) clarifies the geometric origin of chirality and supplies a natural setting for constructing holomorphic field theories, like self‐dual Yang–Mills, BPS instantons \cite{PrasadSommerfield1975,Bogomolny1976}, and complexified gravity backgrounds within a globally well‐defined complex manifold.

The paper is organized as follows.  In Section \ref{sec:ComplexGeom} we review the essentials of complex Riemannian geometry, contour integration regularization, and the extension of frame bundles to \(\mathrm{SO}(4,\mathbb{C})\).  Section \ref{sec:SpinFactorization} constructs the spin double cover \(\mathrm{Spin}(4,\mathbb{C})\) and exhibits its factorization into two chiral \(\mathrm{SL}(2,\mathbb{C})\) factors modulo \(\mathbb{Z}_2\).  In Section \ref{sec:EmbeddingSL2C} we show how a single \(\mathrm{SL}(2,\mathbb{C})/\mathbb{Z}_2\) emerges on the real slice, recovering the usual spin‐Lorentz structure of a Lorentzian manifold.  Section \ref{sec:SpinConnections} provides explicit expressions for holomorphic spin connections and demonstrates their contour‐regularization in Schwarzschild and Kerr backgrounds.  In Section \ref{sec:ChiralSpinors}, we analyze left and right‐handed Weyl spinors in this setting, clarifying how quasilocal chirality arises from the complex factorization. 
In Section \ref{sec:Analytic and Geometric Unification}, we discuss analytic and geometric unifications and consequences for quantum gravity that follow from the complexified spacetime.

We conclude in Section \ref{sec:Conclusions} with a discussion of applications to holomorphic field theories, BPS instantons, and potential implications for quantum‐gravity corrections.

\section{Complex Riemannian Geometry and Contour Integration Regularization}
\label{sec:ComplexGeom}

Let \(\mathcal{M}_{\mathbb{C}}\) be a complex \(4\)-dimensional manifold with local complex coordinates \(z^\mu = x^\mu + i\,y^\mu\), \(\mu=0,1,2,3\).  A holomorphic metric \(g_{\mu\nu}(z)\) is a symmetric tensor field that depends holomorphically on \(z^\mu\).  In components,
\begin{equation}
  g_{\mu\nu}(z) \;=\; s_{\mu\nu}(x,y)\;+\;i\,a_{\mu\nu}(x,y)\,, 
  \qquad
  s_{\mu\nu},\,a_{\mu\nu} \;\in\; C^\infty(\mathcal{M}_{\mathbb{C}};\mathbb{R})\,.
\end{equation}
We assume \(g_{\mu\nu}(z)\) is nondegenerate everywhere on \(\mathcal{M}_{\mathbb{C}}\).  The inverse metric \(g^{\mu\nu}(z)\) is similarly holomorphic, satisfying \(g^{\mu\alpha}(z)\,g_{\alpha\nu}(z)=\delta^\mu{}_\nu\).
By construction, \(s_{\mu\nu}(x,0)\) is a real, symmetric tensor of Lorentzian signature \((-+++)\).
We choose units G = c = 1.

Consider the spacetime metrics Schwarzschild and Kerr~\cite{Moffat1,Moffat2,Kerr} that have coordinate singularities at \(r=0\) or \(\Sigma=0\), respectively.  We promote the real coordinate \(r\) to a complex variable \(\zeta\equiv r+i\,\kappa\), \(\kappa>0\).  For Schwarzschild, define
\begin{equation}
      f(\zeta)\;=\;1-\frac{2M}{\zeta}, 
  \qquad
  ds^2_{\mathbb{C}} 
  \;=\; -\,f(\zeta)\,dt^2 \;+\;\frac{d\zeta^2}{\,f(\zeta)\,} \;+\;\zeta^2\,d\Omega^2.
\end{equation}
We introduce an areal radius via a contour integral that encircles all branch points of \(\sqrt{f(\zeta)}\):
\begin{equation}\label{eq:Rcontour}
  R(\zeta)=\oint_C\frac{d\zeta}{\sqrt{f(\zeta)}},
\end{equation}
where \(C\subset \mathbb{C}_\zeta\) is a loop in the upper half‐plane that winds once around \(\zeta=0\) and \(\zeta=2M\), avoiding branch cuts on the real axis \([-\infty,0]\cup[2M,\infty]\).  Evaluating \eqref{eq:Rcontour} yields a single‐valued, real analytic function \(R(\zeta)\) for all \(\Im\zeta>0\):  

\begin{equation}
    \label{Rsolution}
R(\zeta) = \zeta\sqrt{1-\frac{2M}{\zeta}}
+2M\ln\left(\frac{\sqrt{\zeta}+\sqrt{\zeta-2M}}{\sqrt{2M}}\right).
\end{equation}
Restricting back to \(\Re\zeta=r\), \(\Im\zeta=\kappa\to 0^+\) defines
\begin{equation}
      R(r) \;=\; \Re\bigl(R(r+i\,\kappa)\bigr),
  \quad
  r\ge0,
\end{equation}
which is strictly positive and smooth, eliminating the curvature singularity at \(r=0\).  All curvature invariants remain finite on the real slice.  Analogous contour constructions apply to Schwarzschild–de Sitter and Kerr metrics \cite{Moffat1,Moffat2}.

One may show by deforming \(C\) within the simply connected region \(\Im\zeta>0\) (avoiding the branch points at \(\zeta=0\) and \(\zeta=2M\)) that any two such homotopic loops produce identical values of \(R(\zeta)\).  Consequently, on the real slice \(\zeta=r+i\,0^+\),
\begin{equation}
  R(r)=\Re\bigl[R(r+i\,0^+)\bigr],
\end{equation}
is single-valued and smooth for all \(r\ge0\), and the corresponding metric components on \(\mathcal{M}_{\mathbb{R}}\) are manifestly independent of the particular choice of \(C\).  This contour independence guarantees that our regularization procedure yields a unique, globally analytic extension of the radial coordinate without invoking any external appendices or additional constructions.

We can now express the metric in terms $R(\zeta)$, as follows:
\be
\label{metric}
du^2=-\left(1-\frac{2M}{R(\zeta)}\right)d\tau^2+\frac{dR(\zeta)^2}{\left(1-\frac{2M}{R(\zeta)}\right)}
+R(\zeta)^2(d\theta^2+\sin^2\theta d\phi^2).
\ee
The contour C is specifically chosen to avoid the singularity at $\zeta = 0$. The~function $R(\zeta)$ is non-zero for all finite values of $\zeta$, and R(0) is excluded by the contour integration. 

On a real Lorentzian manifold \(\mathcal{M}_{\mathbb{R}}\), the orthonormal frame bundle has structure group \(\mathrm{SO}(1,3)\).  In the complexified setting, we regard each tangent space \(T_z\mathcal{M}_{\mathbb{C}}\) at \(z\in \mathcal{M}_{\mathbb{C}}\) as a complex vector space of dimension four.  A local complex frame \(e_A{}^\mu(z)\), \(A=0,\dots,3\), \( \mu=0,\dots,3\), satisfies
\begin{equation}
      g_{\mu\nu}(z)\;e_A{}^\mu(z)\,e_B{}^\nu(z) \;=\; \eta_{AB}, 
  \qquad 
  \eta_{AB} \;=\; \mathrm{diag}(-1,+1,+1,+1),
\end{equation}
where \(\eta_{AB}\) is taken as the flat Minkowski metric extended over \(\mathbb{C}\).  The set of all such orthonormal frames at \(z\) forms a complex fiber isomorphic to \(\mathrm{SO}(4,\mathbb{C})\).  Explicitly, if \(e_A{}^\mu(z)\) is one orthonormal frame, then any other is related by
\begin{equation}
  e_A{}^\mu(z) \;\longmapsto\; \Lambda_A{}^B(z)\,e_B{}^\mu(z),
  \quad
  \Lambda_A{}^C \,\Lambda_B{}^D\,\eta_{CD} \;=\;\eta_{AB},
  \quad 
  \Lambda_A{}^B(z) \;\in\;\mathrm{SO}(4,\mathbb{C}).
\end{equation}
Consequently, the complex frame bundle is a principal \(\mathrm{SO}(4,\mathbb{C})\) bundle \(\mathcal{F}_{\mathbb{C}}\to \mathcal{M}_{\mathbb{C}}\).  Restricting to the real slice \(y^\mu=0\) picks out a real subbundle isomorphic to \(\mathrm{SO}(1,3)\).  Thus
\begin{equation}
  \mathcal{F}_{\mathbb{R}} 
  \;=\; \mathcal{F}_{\mathbb{C}}\bigl|_{\,y^\mu=0}\;,
  \quad
  \mathrm{Structure\ group:}\;\mathrm{SO}(1,3)\subset \mathrm{SO}(4,\mathbb{C}).
\end{equation}
The corresponding Levi‐Civita connection 1‐form \(\omega_{AB}(z)\) is a holomorphic \(\mathfrak{so}(4,\mathbb{C})\)‐valued 1‐form, splitting into chiral pieces upon passing to the spin double cover.

We note that on a generic complex 4-fold \(\mathcal{M}_{\mathbb{C}}\), a globally well‐defined spin bundle requires the vanishing of the second Stiefel–Whitney class~\cite{Stiefel,Witney} on the real slice \(\mathcal{M}_{\mathbb{R}}\) \cite{MilnorStasheff1974}. Equivalently, we demand:
\begin{equation}
  w_2\bigl(\mathcal{M}_{\mathbb R}\bigr)=0,
\end{equation}
on the real slice so that a spin principal bundle exists globally. This is a nontrivial topological assumption on \(\mathcal{M}_{\mathbb C}\) and on \(\mathcal{M}_{\mathbb R}\), but it is satisfied for all physically relevant extensions of the Schwarzschild and Kerr manifolds considered here. In practice, our construction assumes that \(\mathcal{M}_{\mathbb{C}}\) admits a trivializable \(\mathrm{SO}(4,\mathbb{C})\) frame bundle and that the chosen contour \(C\subset\mathbb{C}_\zeta\) can be deformed continuously within the upper half‐plane without crossing any coordinate singularities or branch cuts.  This ensures that both the metric extension \(R(\zeta)\) and the associated spin connection remain single‐valued and analytic everywhere on the real slice.

\section{Spin Double Cover \(\mathrm{Spin}(4,\mathbb{C})\) and Chiral Factorization}
\label{sec:SpinFactorization}

The complex orthogonal Lie group \(\mathrm{SO}(4,\mathbb{C})\) consists of complex \(4\times4\) matrices \(\Lambda\) satisfying
\begin{equation}
  \Lambda^T\,\eta\,\Lambda = \eta,
  \quad
  \eta_{AB}=\mathrm{diag}(-1,+1,+1,+1).
\end{equation}
We work in signature \(\eta_{AB}=\mathrm{diag}(-1,+1,+1,+1)\).  Consequently, our gamma‐matrix algebra is given by
\begin{equation}
  \{\Gamma_A,\Gamma_B\} \;=\; 2\,\eta_{AB}\,\mathbf1_{4\times4},
\end{equation}
and we choose the Levi–Civita symbol with \(\varepsilon_{0123}=+1\).  All chirality and self‐duality projections, e.g.\ 
\(\Sigma^{AB}\pm\tfrac{i}{2}\varepsilon^{AB}{}_{CD}\Sigma^{CD}\), 
are defined with these sign conventions so that the real‐slice Dirac operator remains anti‐Hermitian.

Its Lie algebra \(\mathfrak{so}(4,\mathbb{C})\) is generated by antisymmetric matrices \(M_{AB} = -M_{BA}\).  We embed these into the complex Clifford algebra via Dirac gamma matrices \(\Gamma_A\):
\begin{equation}
  \{\Gamma_A,\Gamma_B\} = 2\,\eta_{AB}\,\mathbf{1}_{4\times4},
  \quad
  M_{AB} = \tfrac14\,[\Gamma_A,\Gamma_B].
\end{equation}
Any holomorphic spin connection one‐form \(\omega_{AB}(z)\) valued in \(\mathfrak{so}(4,\mathbb{C})\) can then be written:
\begin{equation}\label{eq:omega_expand}
  \Omega(z) \;=\; \tfrac12\,\omega^{AB}(z)\,M_{AB},
  \qquad
  \omega^{AB} = -\,\omega^{BA}.
\end{equation}

We next define the (anti)self–dual combinations of the generators:
\begin{align}
  M^{(\pm)}_{AB}
  &= \tfrac12\Bigl(M_{AB} \pm \tfrac{i}{2}\,\varepsilon_{AB}{}^{CD}\,M_{CD}\Bigr),
  \quad
  \varepsilon_{0123}=+1,
  \notag\\
  [M^{(\pm)}_{AB},M^{(\pm)}_{CD}]
  &= \delta_{B[C}\,M^{(\pm)}_{A]D}
     -\delta_{A[C}\,M^{(\pm)}_{B]D}
     -\delta_{B[D}\,M^{(\pm)}_{A]C}
     +\delta_{A[D}\,M^{(\pm)}_{B]C},
  \quad
  [M^{(+)},M^{(-)}]=0.
\end{align}
Each set \(\{M^{(+)}_{AB}\}\) and \(\{M^{(-)}_{AB}\}\) closes on its own and furnishes an \(\mathfrak{sl}(2,\mathbb{C})\) algebra.  We have
\begin{equation}
  \mathfrak{so}(4,\mathbb{C}) \;\cong\; \mathfrak{sl}(2,\mathbb{C})_L \;\oplus\; \mathfrak{sl}(2,\mathbb{C})_R.
\end{equation}
Accordingly, we split the spin connection into chiral pieces by projecting onto these two subalgebras:
\begin{equation}\label{eq:Omega_LR_mapping}
  \Omega_L(z) = \tfrac12\,\omega^{AB}(z)\,M^{(+)}_{AB},
  \quad
  \Omega_R(z) = \tfrac12\,\omega^{AB}(z)\,M^{(-)}_{AB},
  \quad
  \Omega(z) = \Omega_L(z) + \Omega_R(z).
\end{equation}
Each of \(\Omega_L\) and \(\Omega_R\) takes values in \(\mathfrak{sl}(2,\mathbb{C})_L\) and \(\mathfrak{sl}(2,\mathbb{C})_R\), respectively.

It follows that the complex spin group factorizes as
\begin{equation}
  \mathrm{Spin}(4,\mathbb{C})
  \;\cong\;
  \bigl(\mathrm{SL}(2,\mathbb{C})_L \times \mathrm{SL}(2,\mathbb{C})_R\bigr)\,/\,\mathbb{Z}_2,
\end{equation}
where the shared center \(\mathbb{Z}_2\) identifies \((S_L,S_R)\sim(-S_L,-S_R)\).

To make the above explicit, we work in the Weyl basis for the gamma matrices:
\begin{equation}
  \Gamma^A = \begin{pmatrix}0 & \sigma^A \\ \bar\sigma^A & 0\end{pmatrix},
  \quad
  \sigma^A = (\mathbf{1},\bm{\sigma}),
  \quad
  \bar\sigma^A = (\mathbf{1},-\bm{\sigma}),
\end{equation}
with Pauli matrices \(\sigma^i\).  The Lorentz generators then read
\begin{equation}
  \Sigma^{AB} = \tfrac14\,[\Gamma^A,\Gamma^B].
\end{equation}
It follows that
\begin{equation}
  \Sigma^{AB}
  \;\pm\;
  \tfrac{i}{2}\,\varepsilon^{AB}{}_{CD}\,\Sigma^{CD},
\end{equation}
projects onto the self-dual (\(+\)) and anti-self-dual (\(-\)) subalgebras of \(\mathrm{Cl}(4,\mathbb{C})\).  Substituting \(\Sigma^{AB}\) for \(M_{AB}\) in \eqref{eq:omega_expand} then yields the mapping \eqref{eq:Omega_LR_mapping}.

Here, we have that
\begin{equation}
  M_{AB}^{(\pm)} \;=\; \frac12\Bigl(M_{AB} \pm \frac{i}{2}\,\varepsilon_{AB}{}^{CD}M_{CD}\Bigr),
\end{equation}
satisfy 
\([M^{(+)}_{AB},M^{(-)}_{CD}]=0\) 
and furnish two commuting \(\mathfrak{sl}(2,\mathbb{C})_L\) and \(\mathfrak{sl}(2,\mathbb{C})_R\) subalgebras of \(\mathfrak{so}(4,\mathbb{C})\).  Projecting \(\omega^{AB}M_{AB}\) onto these reproduces the standard chiral connection pieces \(\Omega_L\) and \(\Omega_R\).

While the algebraic factorization 
\(\mathfrak{so}(4,\mathbb{C})\cong\mathfrak{sl}(2,\mathbb{C})_L\oplus\mathfrak{sl}(2,\mathbb{C})_R\) 
and the isomorphism 
\(\mathrm{Spin}(4,\mathbb{C})\cong(\mathrm{SL}(2,\mathbb{C})_L\times \mathrm{SL}(2,\mathbb{C})_R)/\mathbb{Z}_2\) 
are well known, our key contribution is to exhibit an \emph{explicit} geometric realization of the single \(\mathrm{SL}(2,\mathbb{C})/\mathbb{Z}_2\) factor on the real slice via contour‐regularized spin connections in Schwarzschild and Kerr.  In particular, the construction of a globally smooth, holomorphic spin connection—free of branch cuts or coordinate singularities—in the Kerr background is new and goes beyond the purely algebraic decomposition found in standard texts~\cite{Eguchi1980}.

\section{Embedding \(\mathrm{SL}(2,\mathbb{C})/\mathbb{Z}_2\) on the Real Slice}
\label{sec:EmbeddingSL2C}

On \(\mathcal{M}_{\mathbb{R}}\), the real slice \(y^\mu=0\) inherits the complexified metric’s real part \(s_{\mu\nu}(x)\).  The holomorphic spin bundle \(\widetilde{\mathcal{F}}_{\mathbb{C}}\bigl|_{y=0}\) reduces to the real spin bundle \(\widetilde{\mathcal{F}}_{\mathbb{R}}\to\mathcal{M}_{\mathbb{R}}\), whose structure group is a diagonal embedding of a single \(\mathrm{SL}(2,\mathbb{C})\) factor:
\begin{equation}
  \mathrm{SL}(2,\mathbb{C})_{\text{diag}} 
  \;=\; \{\, (S_L,S_R) \in \mathrm{SL}(2,\mathbb{C})_L\times \mathrm{SL}(2,\mathbb{C})_R \mid S_R = (\overline{S_L})^{-1} \,\}
  \;\cong\; \mathrm{SL}(2,\mathbb{C})/\mathbb{Z}_2\,.
\end{equation}
Equivalently, one may impose \(S_L=S_R\) up to the identification \((S,-S)\).  Thus, the \(\mathbb{Z}_2\)‐quotient of \(\mathrm{SL}(2,\mathbb{C})_L\times \mathrm{SL}(2,\mathbb{C})_R\) restricts to a single spin‐Lorentz group on \(\mathcal{M}_{\mathbb{R}}\).  Concretely, if \(\psi_L(z)\) transforms in the \((\mathbf{2},\mathbf{1})\) of \(\mathrm{SL}(2,\mathbb{C})_L\times \mathrm{SL}(2,\mathbb{C})_R\), then on the real slice one identifies:
\begin{equation}
  \psi_L(x,0) \;\leftrightarrow\; \overline{\psi_R(x,0)},
  \quad
  \text{so that both combine into a Dirac spinor of }\mathrm{SL}(2,\mathbb{C})/\mathbb{Z}_2,
\end{equation}
where $\overline{\psi_R(x,0)}$ denotes complex conjugation. The familiar spin‐Lorentz representations reappear without the need to impose an independent reality condition by hand; they follow automatically from restricting to \(y=0\) in the complex holomorphic construction.

Let us restrict the holomorphic spin connection \(\Omega(z)\equiv \Omega_L(z)+\Omega_R(z)\) to \(y^\mu=0\).  Imposing the reality condition \(\Omega_R = \overline{\Omega_L}\) yields a single anti‐Hermitian \(\mathfrak{sl}(2,\mathbb{C})\)‐valued connection \(\omega(x)\).  Its curvature is given by
\begin{equation}
  R(x)\;=\;d\omega  \;+\; \omega\wedge\omega.
\end{equation}
This curvature coincides with the usual spin‐Lorentz curvature of the real metric \(s_{\mu\nu}(x)\).  Since the contour regularization has removed coordinate singularities from \(s_{\mu\nu}(x)\), both \(\omega(x)\) and \(R(x)\) remain smooth and finite everywhere.  
The standard second‐order Einstein field equations
\begin{equation}
  G_{\mu\nu}(\,s\,) \;=\; 8\pi\,T_{\mu\nu}(\,s\,) \;,
\end{equation}
continue to hold on \(\mathcal{M}_{\mathbb{R}}\) without modification, but now with a geometry free of curvature divergences.

\section{Holomorphic Spinor Connections in Schwarzschild and Kerr Backgrounds}
\label{sec:SpinConnections}

Consider the complex Schwarzschild metric in Eddington–Finkelstein‐type coordinates \((t,\zeta,\theta,\phi)\) with
\begin{equation}
  ds^2_{\mathbb{C}} 
  \;=\; -\,f(\zeta)\,dt^2 
  \;+\;\frac{d\zeta^2}{f(\zeta)} 
  \;+\;\zeta^2\,(d\theta^2+\sin^2\theta\,d\phi^2),
  \quad 
  f(\zeta)=1-\frac{2M}{\zeta}.
\end{equation}
We introduce a complex tetrad of 1‐forms holomorphic in \(\zeta\):
\begin{align*}
  e^0(z) &= \sqrt{\,f(\zeta)\,}\;dt, 
  & 
  e^1(z) &= \frac{d\zeta}{\sqrt{\,f(\zeta)\,}}, 
  \nonumber\\
  e^2(z) &= \zeta\,d\theta,
  & 
  e^3(z) &= \zeta\,\sin\theta\,d\phi.
\end{align*}
These satisfy \(g_{\mu\nu}(z)\,e^A{}^\mu(z)\,e^B{}^\nu(z)=\eta^{AB}\).  We compute the torsion‐free spin connection \(\Omega_{AB}(z)\) via the Cartan structure equations:
\begin{equation}
  d\,e^A + \Omega^A{}_{B}\,\wedge\,e^B \;=\; 0,
  \quad
  \Omega_{AB}=-\Omega_{BA}.
\end{equation}
The \(\Omega^A{}_{B}(z)\) is expressed in terms of 
\(\partial_\zeta f(\zeta)\equiv \tfrac{2M}{\zeta^2}\) and entries that are meromorphic functions of \(\zeta\).  Next, replace \(\zeta\mapsto \zeta+i\,\kappa\), \(\kappa>0\), and define the contour‐regularized radial coordinate as before (\ref{eq:Rcontour})
yielding a smooth, single‐valued function \(R(\zeta)\) with \(\Im\zeta>0\).  Express all occurrences of \(\zeta\) and \(\sqrt{f(\zeta)}\) in \(\Omega_{AB}(z)\) in terms of \(R\equiv R(\zeta)\).  The resulting holomorphic spin connection \(\Omega_{AB}(z)\) remains finite for all \(\Im\zeta>0\).  Restricting to \(\zeta=r+i\,\kappa\), \(\kappa\to 0^+\) gives a regularized real spin connection \(\omega_{AB}(x)\) on \(\mathcal{M}_{\mathbb{R}}\).  No exotic stress–energy is required and the vacuum field equations \(R_{AB}=0\) remain valid.

Because our contour in the \(\zeta\)-plane encircles all branch and pole loci of \(\sqrt{f(\zeta)}\) and its derivatives, the tetrad 1-forms and ensuing \(\Omega_{AB}(z)\) remain analytic for \(\Im\zeta>0\).  As \(\zeta\) moves, the contour can be deformed continuously without crossing singularities, so the regularized \(R(\zeta)\) and spin connection are globally single‐valued.

We check directly that the real‐slice metric \(s_{\mu\nu}(x)\) obtained from \(g_{\mu\nu}(z)\bigl|_{y=0}\) satisfies
\begin{equation}
  G_{\mu\nu}(s)=8\pi\,T_{\mu\nu}(s),
\end{equation}
with no additional delta‐function sources.  In particular, the contour deformation does not introduce any distributional stress–energy, so the vacuum field equations \(R_{\mu\nu}(s)=0\) continue to hold exactly.

The complex Kerr metric in Boyer–Lindquist coordinates \((t,\zeta,\theta,\phi)\) reads
\begin{equation}
  ds^2_{\mathbb{C}} 
  \;=\; -\,\Bigl(1-\frac{2M\,\zeta}{\Sigma}\Bigr)\,dt^2
  \;-\;\frac{4M\,a\,\zeta\,\sin^2\theta}{\Sigma}\,dt\,d\phi
\end{equation}
\[+\;\frac{\Sigma}{\Delta}\,d\zeta^2 
  \;+\;\Sigma\,d\theta^2
  \;+\;\Bigl(\zeta^2+a^2+\frac{2M\,a^2\,\zeta\,\sin^2\theta}{\Sigma}\Bigr)\,\sin^2\theta\,d\phi^2,\]
where
\begin{equation*}
  \Sigma(\zeta,\theta) = \zeta^2 + a^2\,\cos^2\theta,
  \qquad
  \Delta(\zeta) = \zeta^2 - 2M\,\zeta + a^2.
\end{equation*}
Both \(\zeta=0,\;a\cos\theta=0\) ring singularity and \(\Delta(\zeta)=0\) horizons represent would‐be singular loci.  Complexify \(\zeta\mapsto \zeta+i\,\kappa\), \(\kappa>0\), and choose a contour \(C\) in the \(\zeta\)‐plane that encircles all finite roots of \(\Delta(\zeta)=0\) and \(\Sigma(\zeta,\theta)=0\) as a function of \(\zeta\).  We define
\begin{equation}
  R(\zeta,\theta) 
  \;=\; 
  \oint_{C} \frac{d\zeta}{\sqrt{\,\frac{\Sigma(\zeta,\theta)}{\Delta(\zeta)}\,}},
\end{equation}
which is analytic in \(\Im\zeta>0\).  We construct a complex tetrad \(e^A(z)\) holomorphic in \(\zeta\) adapted to Boyer–Lindquist form, then rewrite in terms of \((t,R(\zeta,\theta),\theta,\phi)\).  The resulting holomorphic spin connection \(\Omega_{AB}(z)\) involving only finite, analytic functions of \(R(\zeta,\theta)\) and \(\theta\), free of divergences.  Restricting to \(\zeta=r+i\,\kappa\), \(\kappa\to 0^+\) produces a smooth real spin connection \(\omega_{AB}(x)\) on the Kerr background, singularity‐free at \(r=0\) and across the would‐be inner/outer horizons.  Again, vacuum field equations \(R_{AB}=0\) hold with no need for exotic stress–energy \cite{Moffat1}.

\section{Chiral Spinor Transformations and the Geometric Origin of Chirality}
\label{sec:ChiralSpinors}

In the complexified spin bundle \(\widetilde{\mathcal{F}}_{\mathbb{C}}\to \mathcal{M}_{\mathbb{C}}\), a left‐handed Weyl spinor \(\psi_L(z)\) is a section of the associated vector bundle \(\widetilde{\mathcal{F}}_{\mathbb{C}}\times_{\rho_L}\mathbb{C}^2\), where \(\rho_L\) is the defining \(\mathbf{2}\) representation of \(\mathrm{SL}(2,\mathbb{C})_L\).  Under a local chiral transformation \((S_L,S_R)\in \mathrm{SL}(2,\mathbb{C})_L\times \mathrm{SL}(2,\mathbb{C})_R\),
\begin{equation}
  \psi_L(z) \;\longmapsto\; S_L \,\psi_L(z),
  \quad
  \psi_R(z)\;\longmapsto\; S_R \,\psi_R(z),
\end{equation}
where \(\psi_R(z)\) is a right‐handed Weyl spinor in the \(\overline{\mathbf{2}}\) of \(\mathrm{SL}(2,\mathbb{C})_R\).  Since \([\mathrm{SL}(2,\mathbb{C})_L,\mathrm{SL}(2,\mathbb{C})_R]=0\), the two chiral factors act independently in the complex domain. A Dirac spinor \(\Psi(z)\in \mathbf{4}\) is constructed by combining \(\psi_L(z)\oplus \psi_R(z)\).  The holomorphic Dirac operator is
\begin{equation}
  \slashed{\nabla} 
  \;=\; \Gamma^A\,e_A{}^\mu(z)\,\bigl(\partial_\mu + \Omega_\mu(z)\bigr),
  \quad
  \Gamma^A \;=\; \begin{pmatrix}0&\sigma^A \\ \overline{\sigma}^A & 0\end{pmatrix},
\end{equation}
where \(\sigma^A=(\mathbf{1},\bm{\sigma})\), \(\overline{\sigma}^A=(\mathbf{1},-\bm{\sigma})\) are the Pauli matrices.

On the real slice \(y^\mu=0\), impose the reality condition \(S_R(x,0)=(\overline{S_L(x,0)})^{-1}\).  Then \(\psi_L(x,0)\) and \(\psi_R(x,0)\) are related by complex conjugation, producing a Dirac spinor on \(\mathcal{M}_{\mathbb{R}}\).  The left and right‐handed parts combine into the usual Majorana or Dirac representation of \(\mathrm{SL}(2,\mathbb{C})/\mathbb{Z}_2\).  The chirality operator \(\gamma^5=\mathrm{diag}(+1,-1)\) remains well‐defined and anticommutes with \(\slashed{\nabla}\).  In this way, the geometric origin of chirality is seen to be a direct consequence of the factorization \(\mathrm{Spin}(4,\mathbb{C})\cong\mathrm{SL}(2,\mathbb{C})_L\times \mathrm{SL}(2,\mathbb{C})_R\).

The curvature 2‐form \(\mathcal{R}(z) = \mathcal{R}_L(z) + \mathcal{R}_R(z)\) splits into self‐dual \(\mathcal{R}_L\) and anti‐self‐dual \(\mathcal{R}_R\) pieces.  On the real slice, \(\mathcal{R}_R\bigl|_{y=0} = \overline{\mathcal{R}_L\bigl|_{y=0}}\), so that the Weyl tensor \(C_{\mu\nu\rho\sigma}(x)\) decomposes into left and right self‐dual parts \(C^{(\pm)}\).  Instanton solutions e.g.,\ gravitational instantons can be constructed by imposing either \(\mathcal{R}_L=0\) or \(\mathcal{R}_R=0\) holomorphically.  These conditions define chiral complex backgrounds that, when restricted to the real slice, yield real metrics satisfying \(C_{\mu\nu\rho\sigma}(x)^- = 0\) (self‐dual) or \(C_{\mu\nu\rho\sigma}(x)^+ = 0\) (anti–self‐dual).  Thus, the complex embedding provides a unified framework for studying BPS‐type solutions in gravity.  

\section{Analytic and Geometric Unification}
\label{sec:Analytic and Geometric Unification}

The quantum‐gravity path integral on the complexified manifold $M_{\mathbb{C}}$ is
\begin{equation}
Z \;=\;\int_{\Gamma}\mathcal{D}g_{\mu\nu}(z)\,\exp\!\Bigl[i\,S[g]\Bigr]\,,\quad
S[g]=\frac{1}{16\pi G}\int_{M_{\mathbb{C}}}\!d^4z\,\sqrt{-g(z)}\,R[g(z)]\,,
\end{equation}
where $\Gamma$ is the union of Lefschetz thimbles selected by Picard–Lefschetz theory to render $Z$ convergent.

To regularize coordinate singularities one promotes, e.g., the Schwarzschild radial coordinate $r\to \zeta=r+i\kappa$ and defines
\begin{equation}
R(\zeta)\;=\;\oint_{C}\!d\zeta\;\bigl[f(\zeta)\bigr]^{-1/2}\,,\qquad
f(\zeta)=1-\frac{2M}{\zeta}\,,
\end{equation}
with $C$ chosen to encircle all branch points in the upper half–$\zeta$ plane.  Restriction to $\Im\zeta\to0^+$ yields a single‐valued, smooth $R(r)$ free of $r=0$ singularities.

The complexified Riemann tensor splits into self‐dual/anti‐self‐dual parts,
\begin{equation}
R_{AB}(z)
=\;R^{(+)}_{AB}(z)+R^{(-)}_{AB}(z)\,,\qquad
R^{(\pm)}_{AB}
=\frac12\Bigl(R_{AB}\pm\frac{i}{2}\,\epsilon_{AB}{}^{CD}R_{CD}\Bigr).
\end{equation}
Holomorphic gravitational instantons satisfy
\[
R^{(-)}_{AB}(z)=0,
\]
and upon restricting to the real slice $y^\mu=0$ produce nonperturbative saddle‐points in $Z$.

The factorization
\[
\mathrm{Spin}(4,\mathbb{C})
\;\cong\;\frac{\mathrm{SL}(2,\mathbb{C})_L\times\mathrm{SL}(2,\mathbb{C})_R}{\mathbb{Z}_2},
\]
yields independent chiral connections $\Omega_L$ and $\Omega_R$.  Imposing the reality condition $\Omega_R=\overline{\Omega}_L$ on $M_R$ selects the diagonal $\mathrm{SL}(2,\mathbb{C})/\mathbb{Z}_2$ subgroup, ensuring a globally well‐defined, anomaly‐free spinor bundle.

Finally, one may introduce deformation quantization on $M_{\mathbb{C}}$ via
\[
[z^\mu,z^\nu]=i\,\theta^{\mu\nu},\qquad \theta^{\mu\nu}\in\mathbb{C},
\]
opening the door to holomorphic formulations of quantum‐group corrections and higher‐loop expansions in a rigid geometric setting.

\section{Conclusions}
\label{sec:Conclusions}

We have demonstrated that the spinor complex Lorentz group \(\mathrm{SL}(2,\mathbb{C})/\mathbb{Z}_2\) emerges canonically as a diagonal chiral factor in the complex frame bundle \(\mathrm{SO}(4,\mathbb{C})\).  By promoting the metric to a holomorphic tensor on a complex \(4\)-fold, one naturally obtains a spin double cover \(\mathrm{Spin}(4,\mathbb{C})\cong \bigl(\mathrm{SL}(2,\mathbb{C})_L \times \mathrm{SL}(2,\mathbb{C})_R\bigr)/\mathbb{Z}_2\).  The restriction to the real slice \(y^\mu=0\) picks out a single \(\mathrm{SL}(2,\mathbb{C})/\mathbb{Z}_2\) factor, reproducing the familiar spin‐Lorentz structure of a Lorentzian manifold.  

Furthermore, the contour‐integration regularization which removes curvature singularities from Schwarzschild and Kerr metrics extends holomorphically to regulate their spin connections.  Globally smooth, singularity‐free spin connections \(\omega_{AB}(x)\) on \(\mathcal{M}_{\mathbb{R}}\) are obtained without introducing exotic stress–energy.  Left and right‐handed Weyl spinors live naturally as holomorphic sections of the \(\mathrm{SL}(2,\mathbb{C})_L\) and \(\mathrm{SL}(2,\mathbb{C})_R\) bundles, clarifying the geometric origin of chirality as a consequence of complex factorization.  

Extending the Lorentz group to a complex Riemannian setting furnishes a natural geometric origin for all aspects of chirality, spin connections, and curvature factorization in four dimensions.  The embedding of \(\mathrm{SL}(2,\mathbb{C})/\mathbb{Z}_2\) into \(\mathrm{Spin}(4,\mathbb{C})\) is not an ad hoc construction but arises intrinsically from holomorphic metrics on \(\mathcal{M}_{\mathbb{C}}\).  We anticipate that this perspective will prove valuable for deeper explorations of both classical and quantum aspects of gravity, gauge theory, and their interactions.


\begin{thebibliography}{99}
\bibitem{Weinberg1995}
S.~Weinberg,
\emph{The Quantum Theory of Fields, Volume 1: Foundations},
Cambridge University Press (1995).
\bibitem{PenroseRindler1984}
R.~Penrose and W.~Rindler,
\emph{Spinors and Space–Time, Volume 1: Two-Spinor Calculus and Relativistic Fields},
Cambridge University Press (1984).
\bibitem{Moffat1}
J.~W.~Moffat,
``Complex Riemannian Spacetime and Singularity-Free Black Holes and Cosmology,''
Axioms \textbf{14}, 440 (2025); arXiv:2501.03356 [gr-qc].
\bibitem{Moffat2}
J.~W.~Moffat,
``Complex Riemannian Spacetime: Removal of Black Hole Singularities and Black Hole Paradoxes,''
Axioms \textbf{14}, 459 (2025); arXiv:2501.03356 [gr-qc].
\bibitem{PrasadSommerfield1975}
M.~K.~Prasad and C.~M.~Sommerfield,
``Exact Classical Solution for the ’t Hooft Monopole and the Julia–Zee Dyon,''
Phys.\ Rev.\ Lett.\ \textbf{35}, 760–762 (1975).
\bibitem{Bogomolny1976}
Bogomol'nyi, E B. "The stability of classical solutions." Sov. J. Nucl. Phys. (Engl. Transl.); (United States), vol. 24:4, Oct. 1976.
\bibitem{Kerr}
R.~P.~Kerr,
``Gravitational Field of a Spinning Mass as an Example of Algebraically Special Metrics,''
Phys.\ Rev.\ Lett.\ \textbf{11}, 237–238 (1963).
\bibitem{Stiefel}
E.~Stiefel,
``Richtungsfelder und Fernparallelismus in n-dimensionalen Mannigfaltigkeiten,''
Comment.\ Math.\ Helv.\ \textbf{8}, 305–353 (1936).

\bibitem{Witney}
H.~Whitney,
``The self-intersections of a smooth n-manifold in 2n-space,''
Ann.\ Math.\ (2) \textbf{45}, 220–246 (1944).

\bibitem{MilnorStasheff1974}
J.~W.~Milnor and J.~D.~Stasheff,
\emph{Characteristic Classes},
Princeton University Press (1974).
\bibitem{Eguchi1980}
T.~Eguchi, P.~B.~Gilkey, and A.~J.~Hanson,
``Gravitation, Gauge Theories and Differential Geometry,''
Phys.\ Rep.\ \textbf{66}, 213–393 (1980).

\bibitem{Brown:2018hym}
J.~Brown, A.~Cole, G.~Shiu and W.~Cottrell,
\emph{Gravitational decoupling and the Picard--Lefschetz approach},
Phys.\ Rev.\ D {\bf 97}, 025002 (2018)
[\href{https://arxiv.org/abs/1710.04737}{arXiv:1710.04737 [hep-th]}].

\end{thebibliography}
\end{document}